\documentclass[a4paper]{article}
\usepackage{amsmath}
\usepackage{amsfonts}
\usepackage{amssymb}

\input{tcilatex}

\begin{document}

\title{\textbf{Bohr-Heisenberg Reality and System-Free Quantum Mechanics}}
\author{\textbf{George Jaroszkiewicz$^{*}$ and Jon Eakins} \\
$^{*}$School of Mathematical Sciences, University of Nottingham\\
University Park, Nottingham NG7 2RD, UK}
\date{\today }
\maketitle

\begin{abstract}
Motivated by Heisenberg's assertion that electron trajectories do not exist
until they are observed, we present a new approach to quantum mechanics in
which the concept of observer independent system under observation is
eliminated. Instead, the focus is only on observers and apparatus, the
former describing the latter in terms of labstates. These are quantum states
over time-dependent Heisenberg nets, which are quantum registers of qubits
representing information gateways accessible to the observers. We discuss
the motivation for this approach and lay down the basic principles and
mathematical notation.
\end{abstract}

\section{Introduction}

System-free quantum mechanics (SFQM) is an alternative and powerful way of
thinking about and describing quantum processes. In principle it should do
everything that standard quantum mechanics (SQM) can do and more; it should
allow a discussion of situations involving more than one observer and
provide a framework for describing dynamically coupled sequences of
experiments, such as networks of Mach-Zender interferometers coupled in
parallel, series, or both. Although our approach is novel, it was motivated
by and is the logical consequence of Heisenberg's view of reality, which led
to matrix mechanics in 1925 \cite{HEISENBERG-1925}. Paradoxically,
Heisenberg's approach to quantum mechanics (QM) appeared too metaphysical
and mathematically challenging for many theorists, and quickly lost ground
to the more intuitive wave-mechanical formalism introduced by Schr\"{o}%
dinger the following year. Although these radically different formulations
were soon shown to be mathematically equivalent by Schr\"{o}dinger himself
\cite{SCHRODINGER-1926}, deep differences remained between the views of
Heisenberg and Schr\"{o}dinger on the nature of physical reality. The former
wished to focus only on what could be observed, whilst the latter believed
in an underlying classical reality generating those observations.

Heisenberg restated his views in his uncertainty paper of $1927$ \cite%
{HEISENBERG-1927}. In the context of the correspondence principle \cite%
{PLOTNITSKY} he wrote: \textquotedblleft I believe that the existence of the
classical `path' can be pregnantly formulated as follows: The `path' comes
into existence only when we observe it.\textquotedblright\ This statement is
the foundation of the system-free approach to QM.

We shall call the principle that \textquotedblleft if we have not observed
something then it does not exist\textquotedblright\ \emph{quantum
counterfactuality}, or \emph{Heisenberg's reality principle}. In SFQM we
take quantum counterfactuality to its logical conclusion and assume that
systems under observation themselves do not exist independently of any
contextual observation. Because Bohr's views on QM were allied to those of
Heisenberg, we shall also refer to this idea as \emph{Bohr-Heisenberg reality%
}.

In the next section we discuss the properties of observers and apparatus,
which are central to our approach. Then we introduce the notion of a \emph{%
Heisenberg net}, which provides the mathematical basis for our description.
A\ Heisenberg net is a representation of all the information gateways
accessible to an observer at a given time. This leads naturally to a
discussion of the \emph{preferred basis} and \emph{signal states}. This puts
us in a position to introduce the central concept of the \emph{labstate},
which encodes the potential outcomes of an experiment in terms of the
signals which the apparatus may emit. Dynamics is discussed in terms of how
the labstate changes in time, requiring us to bring in the novel concepts of
\emph{Born maps, semi-unitary operators }and\emph{\ Schr\"{o}dinger evolution%
}. Finally, we discuss how signal operators evolve and give an expression
for path summations, which are the SFQM analogues of Feynman path integrals
in SQM. Nowhere do we deal with a state of a system under observation.

\section{Observers}

Following Heisenberg, the objective of SFQM is to focus as much as possible
on those aspects of quantum physics which are physically meaningful, so what
can we be sure of? When we describe a physics experiment, we can always be
sure of two things: first, that there is an \emph{observer} and second, that
they use some \emph{apparatus}. In SFQM, observers and apparatus are primary
concepts which have to be taken as given. However, we can say some things
about their properties.

In SFQM, an observer is always a physically real, classically autonomous
subset of the universe with a sense of time, which can manipulate or
interact with another physically real subset of the universe called the
apparatus. Each observer has a time-dependent \emph{information/memory
content} and a set of \emph{rules} for processing this information and
manipulating their apparatus accordingly. There is no necessity to regard an
observer as conscious, or alive in a biological sense.

The time associated with any observer is always discrete, but this
discreteness is not necessarily defined in terms of equal standard intervals
of time. Temporal discreteness arises because information is gained or lost
by an observer only in stages, such as during state preparation or quantum
outcome detection. Now from the point of view of decoherence theory or
quantum field theory, it could be argued that we cannot always be absolutely
certain when information has been extracted. In practice however,
experimentalists really have no such problem, and it is this fact which we
rely on here. Everything that experimentalists ever do amounts to the
collection of answers to elementary \emph{yes/no} questions. Moreover,
although ordinary experience leads us to imagine that time is continuous,
there is actually no hard evidence for that view. On the contrary, all of
quantum physics is based on a comparison between initial and final states,
which necessarily introduces a discrete view of time.

In SFQM, it is possible to discuss as many different observers as we want.
Each observer will have their own sense of time, relative to which both
their associated apparatus and information content may change. An important
rule which is an essential and established feature of quantum
counterfactuality is that \emph{different observers never share apparatus or
information content} at the same time. If they did, they would have to be
regarded as a single observer with a single apparatus. This has significant
implications for relativistic physics, which we do not have space to comment
on here.

\section{Heisenberg nets}

Although SFQM is a consistent quantum theory fully backwards compatible with
SQM, it is sufficiently different in its core values to warrant a modified
notation for state vectors and operators. Recall that in SQM, states are
often described by Dirac kets with angular brackets, such as $|\psi\rangle$.
In SFQM we replace the angular brackets by round brackets, such as $|\psi) $%
. Then it will always be clear which formalism is being used. This is very
important, because conceptually quite different Hilbert spaces are involved
in describing what looks superficially like the same thing. As for
operators, we shall always denote SFQM operators acting over Heisenberg nets
in blackboard bold font, such as $\mathbb{A}_{i}^{+}$.

We provide an explicit mathematical representation of SFQM using the concept
of a \emph{Heisenberg net}, defined as follows.

First, we use the fact that an observer comes provided with a sense of time.
This means that for a given observer, there is a notion of simultaneity.
Different parts of a laboratory will be described at the same instants of
the observer's time, very much as if there was some sort of absolute time.
Different observers need have no sense of a common time or rate of time,
however. That is consistent with relativity, which emphasizes the physical
significance of non-integrable proper time compared to the purely formal
integrable coordinate time used to discuss points in the spacetime manifold.
Whenever we discuss a Heisenberg net, we will be thinking about it at a
given instant of time in some local laboratory frame of reference. This
frame need not be regarded an inertial frame, or even a freely-falling
laboratory, so in this respect we anticipate the eventual construction of a
SFQM approach to general relativity.

Second, at a given instant $n$ of an observer's time, the observer assigns
one quantum bit (qubit) to each information gateway, i.e., each piece of
their apparatus where information could in principle be inserted into, or
extracted from, the current state of their apparatus by them at that time.
What such a qubit means will be further explained below.

The Heisenberg net $\mathcal{H}_{n}$ associated with a given observer at
time $n$ is just the Hilbert space of all those qubits associated with their
apparatus at that time. In SFQM, we shall assume that apparatus may change
dynamically, so the associated Heisenberg net changes in consequence. At any
given time we shall always suppose that there is a finite number, $r_{n}$,
of such qubits. This is the only assumption that makes sense in the context
of our description, because \emph{there are no real quantum experiments
which can extract an infinite amount of information from any source}. We
note here a comment by Marvin Minsky on Feynman's ideas on the
representation of physics as a computational process \cite{FEYNMAN-1982}.
Minsky wrote that Feynman did not like the concept of continuous space,
because he could not see how a finite volume of space could contain an
infinite amount of information \cite{MINSKY-2000}.

We label the qubits in the Heisenberg net at time $n$ by $\mathcal{Q}%
_{n}^{1} $, $\mathcal{Q}_{n}^{2},\ldots,\mathcal{Q}_{n}^{r_{n}}.$ How the
upper index is assigned is arbitrary, there being no requirement at this
stage to think of some of the qubits as nearest neighbours, or even
ancestors of earlier qubits with the same indices. These qubits form a
quantum register, or Heisenberg net $\mathcal{H}_{n}$, which is the tensor
product $\mathcal{H}_{n}\equiv \mathcal{Q}_{n}^{1}\otimes\mathcal{Q}%
_{n}^{2}\otimes\ldots\otimes \mathcal{Q}_{n}^{r_{n}}$. This is a Hilbert
space of dimension $d_{n} \equiv2^{r_{n}}$. The number $r_{n}$ of qubits
will be referred to as the \emph{rank} of the Heisenberg net at time $n$.

Before we discuss what states in a Heisenberg net register represent, let us
clarify what the qubits making up the register represent. Most importantly
and contrary to what might be expected, a given SFQM qubit does not
represent the two distinct physical outcomes, such as spin \emph{up} or spin
\emph{down}, of experiments such as the Stern-Gerlach experiment. Rather,
each qubit represents a separate outcome detector, or detector site, or
whatever constitutes a signal detector, associated with a single outcome.
The two possible outcomes of an idealized Stern-Gerlach apparatus therefore
require two SFQM qubits: one for the spin-up outcome and another for the
spin-down outcome. Likewise, any apparatus described according to SQM with $%
k $ possible outcomes would require $k$ qubits. This applies to situations
where the PVM (projection valued measure) or the POVM (positive operator
valued measure) formulations in SQM are used. The rule is simply to look at
the laboratory, determine all those part of the apparatus where information
could be obtained in elementary yes/no terms, and assign a qubit to each one.

Each Heisenberg net qubit has two possible physically observable states, but
only one of these represents a positive physical signal in the associated
detector. One of these states, denoted by $|0)$, represents the detector in
the \emph{void} state. If the observer looked at the detector when it was in
that state, it would register nothing, i.e., no signal. The other possible
state of the detector is denoted by $|1)$ and represents the fact that,
\emph{if} the observer looked at that detector when it was in that state, it
would register a signal.

We emphasize several points. First, these detectors are not necessarily
localized in space. For example, some of them could be associated with
momentum, and as such, would not necessarily have a spatially localized
physical realization. Each Heisenberg net qubit represents an elementary
information gateway for the potential modification of the information
content held by the observer, and this is rather general. This information
is an essential ingredient in SFQM in another way; it is used by the
observer to interpret what each qubit means. Without such information,
apparatus is meaningless.

Second, SFQM is a universal theory, in that in principle its
formalism should be applicable to any quantum experiment. We do
not need to think only in terms of electrons and photons. What
distinguishes one type of experiment from another, even in those
case where the Heisenberg nets appear identical, will be the
information content and the subsequent dynamical evolution.

Third, quantum counterfactuality is used throughout. We are dealing with a
quantum theory, and therefore, the possible states of a Heisenberg net,
referred to as \emph{labstates}, represent a form of potentiality for
observation in the future. This means that, if the observer chooses not to
look at their detectors at time $n$ but "passes the labstate on to new
apparatus", then the labstate at that time becomes the initial labstate for
the next jump forwards in time. The formalism requires us in practice to
restrict the discussion of all quantum processes to the forwards direction
of time only.

For a given detector/information gateway $P$, the associated qubit $\mathcal{%
Q}_{P}$ has all of the structure of a two dimensional Hilbert space. Given
the natural basis states $\{|0)_{P},|1)_{P}\}$ for $\mathcal{Q}_{P}$, we
define the signal excitation operator $a_{P}^{+}\equiv|1)_{P}(0|$ and its
dual, $a_{P}\equiv|0)_{P}(1|$. Apart from the identity operator $I_{P}$,
these are the only qubit operators we shall use.

\section{The preferred basis}

SFQM avoids the well-known preferred basis problem which haunts SQM because
there is a natural mechanism for selecting such a basis. This is the
information content $I_{n}$ held by the observer at time $n$. Given any
piece of detecting equipment, the observer will always know what amounts to
a lack of a signal and what represents a signal. Given this knowledge, it is
a trivial matter to associate the corresponding basis vectors $|0)$ and $|1)$
with this information for each qubit. Extending this to all the qubits in
the current Heisenberg net gives $B_{n},$ the natural, preferred basis for $%
\mathcal{H}_{n}$.

There are various equivalent representations of $B_{n}$ which we shall use
as circumstances dictate. The \emph{product representation} describes the $%
d_{n}\equiv2^{r_{n}}$ elements of $B_{n}$ in terms of tensor products of
individual qubit states, i.e.,%
\begin{align}
B_{n} &
=\{|0,n)_{1}|0,n)_{2}\ldots|0,n)_{r_{n}},|1,n)_{1}|0,n)_{2}%
\ldots|0,n)_{r_{n}},\ldots  \notag \\
& \ \ \ \ \ \ \ \ \ \ \ \ \ \ \ \ \ \ \ \
\ldots,|1,n)_{1}|1,n)_{2}\ldots|1,n)_{r_{n}}\},
\end{align}
where for example $|i,n)_{j}$ represents state $i$ of the $j^{th}$ qubit in
the Heisenberg net at time $n$. Here we have suppressed the tensor product
symbol.

The \emph{occupation representation} is a more compact version of the above.
Now we label the $d_{n}$ elements of $B_{n}$ in terms of the associated
finite binary sequences, i.e.,
\begin{equation}
B_{n}=\{|00\ldots0,n),|10\ldots0,n),\ldots,|11\ldots1,n)\}.
\end{equation}

The \emph{signal representation }is based on excitations of the \emph{void
state }$|0,n)\equiv|00\ldots0,n)$. First, we define the signal operators $%
\left\{ \mathbb{A}_{i,n}^{+}:i=1,2,\ldots,r_{n}\right\} $ by
\begin{equation}
\mathbb{A}_{i,n}^{+}\equiv I_{1,n} I_{2,n} \ldots I_{i-1,n} a_{i,n}^{+}
I_{i+1,n} \ldots I_{r_{n},n},\ \ \ 1\leqslant i\leqslant r_{n},
\end{equation}
where again we have suppressed the tensor product symbol, and then we may
write%
\begin{equation}
B_{n}=\left\{ |0,n),\mathbb{A}_{1,n}^{+}|0,n),\mathbb{A}_{2,n}^{+}|0,n),%
\ldots,\mathbb{A}_{1,n}^{+}\mathbb{A}_{2,n}^{+}\ldots\mathbb{A}%
_{r_{n},n}^{+}|0,n)\right\} .
\end{equation}

The signal representation is particularly well-suited for SFQM. The basis
set can be partitioned into disjoint signal classes, defined by the number
of signal operators involved. The zero-signal class consists of just one
element, the void state. The one-signal class consists of basis states of
the form $\mathbb{A}_{i,n}^{+}|0,n)$, and there will be $r_{n}$ such states,
and so on. More generally, the $k-$signal class consists of states of the
form $\mathbb{A}_{i_{1},n}^{+}\mathbb{A}_{i_{2},n}^{+}\ldots\mathbb{A}%
_{i_{k},n}^{+}|0,n)$ and there are $\binom{r_{n}}{k}$ such elements in $%
B_{n} $. Counting up all the elements in all the signal classes gives us a
total of $2^{r_{n}}$, as expected.

An arbitrary \emph{one-signal labstate} will be a superposition of
one-signal basis states, i.e.,%
\begin{equation}
|\psi,n)=\sum_{i=1}^{r_{n}}\psi_{n}^{i}\mathbb{A}_{i,n}^{+}|0,n),\ \ \ \psi
_{n}^{i}\in\mathbb{C},
\end{equation}
suitably normalized to unity. For such a labstate, any actual outcome would
involve a signal from only one detector site, with all the others remaining
void. Quantum uncertainty means that, before the observer looked, they would
not in general know for sure which one of the detectors would fire.
Non-locality occurs here in terms of the apparatus detectors, not in terms
of any system under observation. This effectively eliminates the
wave-particle duality issue.

One-signal labstates model those situations when we know we are dealing with
a one-particle system, such as a single electron or single photon, for
example. Many traditional scenarios in SQM can be discussed solely in terms
of such states, but we note that SFQM goes beyond one-signal labstates. For
example, experiments involving pairs of photons in the initial state would
be described by two-signal labstates in SFQM \cite{JAROSZKIEWICZ-2006}.
Experiments such as those discussed by Einstein, Podolsky and Rosen (EPR)
\cite{EPR-1935}, which involve two separated detectors firing
simultaneously, would also require the use of two-signal labstates. There is
nothing in the formalism which forbids superposition of different signal
class elements, or dynamical changes of signal class. In this respect, SFQM
is more like quantum field theory rather than Schr\"{o}dinger wave mechanics.

The \emph{computation representation} is based on the fact that any finite
binary sequence $\{\varepsilon_{1},\varepsilon_{2},\ldots,\varepsilon
_{r}\},\varepsilon_{i}=0$ or$\ 1,$ can be mapped into a unique integer $i$
in the range $0\leqslant i<2^{r}$ via the computational map $i=\varepsilon
_{1}2^{0}+\varepsilon_{2}2^{1}+\ldots+\varepsilon_{r}2^{r-1}$. Then we may
write%
\begin{equation}
B_{n}=\left\{ |i,n):i=0,1,2, \ldots,2^{r_{n}}-1\right\} .
\end{equation}
Orthonormality is best described in terms of this representation, because
now we can write $(i,n|j,n)=\delta_{ij}$, $0\leqslant i,j<2^{r_{n}}$.
Completeness and the resolution of the identity $\mathbb{I}_{n}$ are also
best described in this representation.

\section{Labstates}

The \emph{labstate} $| \psi,n)$ is the current quantum state of an
observer's Heisenberg net at time $n$. We shall restrict our attention in
this paper to pure labstates, which means that Heisenberg nets are regarded
as certain and labstates are single elements in them. We have no space here
to discuss mixed labstates, which will occur in situations where the
observer's information about the current labstate or the Heisenberg net
itself is incomplete.

In general, labstates will always be normalized to unity, because the
standard Born probability rules are assumed and there is no concept here of
leakage of probability. If for example a signal corresponding to a particle
"disappears", the probability associated with it is transferred to the void
state. For a given labstate, the probability that the apparatus is void
added to the sum of the probabilities that one or more detectors have fired
always sums up to unity. Using the computation representation of the basis $%
B_{n}$, a labstate can always be written in the form
\begin{equation}
|\psi,n)=\sum_{i=0}^{{d}_{n}-1}\psi_{n}^{i}|i,n),\ \ \ \ \ \ \ \ \ \ \ \sum
_{i=0}^{{d}_{n}-1}|\psi_{n}^{i}|^{2}=1.
\end{equation}
The coefficients $\left\{ \psi_{n}^{i}\right\} $ in this expansion have the
usual Born probability interpretation: $|\psi_{n}^{i}|^{2}$ is the
probability that, given that particular labstate at time $n$, all of the
detectors associated with the preferred basis vector $|i,n)$ would each be
in their signal state. For a given integer $i$ in the range $[1,{d}_{n}-1]$,
we can determine which qubits are in their signal state by inverting the
computational map and decomposing $i$ as a unique sum of terms in the form $%
i=2^{j_{1}-1}+2^{j_{2}-1}+\ldots+2^{j_{k}-1}$, where $j_{1}<j_{2}<%
\ldots<j_{k}$ are non-negative integers, for some non-zero $k$. Then we can
write
\begin{equation}
|i,n)=\mathbb{A}_{j_{1},n}^{+}\mathbb{A}_{j_{2},n}^{+}\ldots\mathbb{A}%
_{j_{k},n}^{+}|0,n),
\end{equation}
for $0<i<d_{n}$, which means that the detectors corresponding to qubits $%
\mathcal{Q}_{n}^{j_{1}},\mathcal{Q}_{n}^{j_{2}},\ldots,\mathcal{Q}%
_{n}^{j_{k}}$ will fire whilst all the others remain void. To illustrate the
point, suppose we have a rank-two Heisenberg net at time $n$. Then the most
general labstate is of the form%
\begin{equation}
|\psi,n)=\left\{ \alpha\mathbb{+}\beta\mathbb{A}_{1,n}^{+}+\gamma \mathbb{A}%
_{2,n}^{+}+\delta\mathbb{A}_{1,n}^{+}\mathbb{A}_{2,n}^{+}\right\} |0,n),
\end{equation}
with $|\alpha|^{2}+|\beta|^{2}+|\gamma|^{2}+|\delta|^{2}=1$. Then the
probability that no signal would fire is $|\alpha|^{2}$, the probability
that only detector one would fire is $|\beta|^{2}$, the probability that
only detector two would fire is $|\gamma|^{2}$, and the probability that
detectors one and two would fire simultaneously is $|\delta|^{2}$, \emph{if}
the observer looked.

\section{Dynamics}

SFQM is designed to encode quantum dynamics in as realistic a way as
possible, which means that there is no concept here of inconsistent
histories or many worlds. An important characteristic of SFQM is that \emph{%
there is no scope for unphysical quantum states to occur.} The only states
considered are labstates, and these are always physically meaningful.

The information content held by an observer at time $n$ is regarded as
classically certain and objective, relative to that observer, but it is not
constant in time. As the observer's time progresses, several things change
with it, including the information content, the labstate, and the Heisenberg
net itself.

\subsection{Born maps}

Given two finite dimensional Hilbert spaces $\mathcal{H}$, $\mathcal{H}%
^{\prime}$, we define a \emph{Born map} $\mathfrak{B}$ to be a
norm-preserving map from $\mathcal{H}$ into $\mathcal{H}^{\prime}$, i.e.,
for any state $\psi\in\mathcal{H}$, the image $\psi^{\prime}\equiv\mathfrak{B%
}\psi$ of $\psi$ is an element of $\mathcal{H}^{\prime}$ such that $%
(\psi^{\prime},\psi^{\prime})=(\psi,\psi)$. Born maps are not necessarily
linear and the two Hilbert spaces involved need not have the same dimension.

In SFQM, Born maps may be used to discuss state preparation and outcome
detection. These represent processes involving non-trivial changes in an
observer's information content. In SQM, such actions are described by
non-unitary evolution and are normally associated with state reduction, or
wave-function collapse. These are commonly regarded as a blemish on an
otherwise beautiful theory. There should not be the same stigma attached to
non-linear Born evolution in SFQM, because the collapse does not refer to
any supposed changes in a system under observation, but to changes in the
information content held by an observer.

\subsection{Semi-unitary operators}

In SQM, the observer's information content does not change between state
preparation and outcome detection. This is most obvious in the Heisenberg
picture. In the formally equivalent Schr\"{o}dinger picture, evolution of
the wavefunction is unitary, which involves a linear operator. In SFQM,
therefore, we shall assume that evolution between state preparation and
outcome detection involves linear Born maps. It turns out that linearity
imposes important restrictions on the evolution operators which have
far-reaching consequences.

We define a \emph{semi-unitary} operator as a linear Born map. It is easy to
prove that if $U$ is a semi-unitary operator from $\mathcal{H}$ to $\mathcal{%
H}^{\prime}$, then $\dim\mathcal{H}\leqslant\dim\mathcal{H}^{\prime }$ and $%
U^{+}U=I$, the identity operator over $\mathcal{H}$. If $\dim \mathcal{H}%
<\dim\mathcal{H}^{\prime}$, then it is easy to see that $UU^{+}\neq
I^{\prime}$, the identity operator over $\mathcal{H}^{\prime}$. Moreover, if
$U$ is semi-unitary, then inner products, and not just norms, are preserved,
i.e., if $\psi^{\prime}\equiv U\psi$, $\phi^{\prime}\equiv U\phi$, then $%
(\psi^{\prime},\phi^{\prime})=(\psi,\phi)$ for any vectors $\psi$, $\phi$ in
$\mathcal{H}$.

We now apply these ideas to SFQM. If the dynamical evolution operator $%
\mathbb{U}_{n+1,n}$ carrying the labstate from $\mathcal{H}_{n}$ to $%
\mathcal{H}_{n+1}$ is semi-unitary then for evolution given by
\begin{equation}
|\psi,n)\rightarrow|{\psi}^{\prime},n+1)\equiv\mathbb{U}_{n+1,n}|\psi,n),
\end{equation}
we know inner products are preserved, i.e.,
\begin{equation}
({\psi}^{\prime},n+1|{\phi}^{\prime},n+1)=(\psi,n|\phi,n),
\end{equation}
for all $|\psi,n),|\phi,n)$ in $\mathcal{H}_{n}$, and from this we can prove
that total probability will be conserved.

We may express semi-unitary evolution of the preferred basis $B_{n}$ in
terms of the preferred basis $B_{n+1}$ at time $n+1$, i.e., writing
\begin{equation}
\mathbb{U}_{n+1,n}|i,n)=\sum_{j=0}^{{d}_{n+1}-1}U_{n+1,n}^{j,i}|j,n+1),
\end{equation}
where the coefficients $\left\{ U_{n+1,n}^{j,i}\right\} $ satisfy the
semi-unitarity equations
\begin{equation}
\sum_{k=0}^{{d}_{n+1}-1}\left( U_{n+1,n}^{k,j}\right)
^{\ast}U_{n+1,n}^{k,i}=\delta_{ij}.
\end{equation}

\subsection{Schr\"{o}dinger evolution}

In SFQM, information held by the observer about these coefficients is
equivalent to knowledge of the Hamiltonian in SQM, which allows the
observer, in that formulation, to make dynamical calculations via the Schr%
\"{o}dinger equation.

Typical quantum experiments are, by construction, \emph{closed}, which means
that, as far as physically possible, all external processes are excluded
from interaction with the apparatus in the time between state preparation
and outcome detection. In SFQM, an important rule characterizing
semi-unitary evolution for such situations which distinguishes it from state
preparation and outcome detection is its action on the void state. Closed
semi-unitary evolution describes the behaviour of laboratory equipment in
the absence of any active intervention by the observer or any external
agency. In such circumstances, it would be bizarre if the void labstate
spontaneously changed into a non-zero signal labstate during a closed
experiment. Likewise, a non-zero signal labstate would normally evolve into
some other non-zero signal labstate in the absence of any active
intervention during a closed experiment

This leads us to formulate the following rule: closed semi-unitary evolution
in SFQM, corresponding to Schr\"{o}dinger (unitary) evolution in SQM, will
generally evolve a void labstate into a void labstate, whilst non-zero
signal labstates evolve into non-zero signal labstates. Expressed
mathematically, this means
\begin{equation}
\mathbb{U}_{n+1,n}|0,n)=|0,n+1)  \label{222}
\end{equation}
for any closed experiment and
\begin{equation}
(0,n+1|\mathbb{U}_{n+1,n}|\psi,n)=0  \label{333}
\end{equation}
for any non-zero signal labstate $|\psi,n)$, which by definition satisfies
the rule $(0,n|\psi,n)=0$. We shall call any evolution satisfying $(\ref{222}%
)$ and $(\ref{333})$ \emph{Schr\"{o}dinger evolution.}

For Schr\"{o}dinger evolution in SFQM, the corresponding matrix $%
[U_{n+1,n}^{i,j}]$ takes the form%
\begin{equation}
\lbrack U_{n+1,n}^{i,j}]=\left[
\begin{array}{llll}
1 & 0 & \ldots & 0 \\
0 & U_{n+1,n}^{1,1} & \ldots & U_{n+1,n}^{1,{d}_{n}-1} \\
\vdots & \vdots &  & \vdots \\
0 & U_{n+1,n}^{{d}_{n+1}-1,1} & \ldots & U_{n+1,n}^{{d}_{n+1}-1,{d}_{n}-1}%
\end{array}
\right] ,\ \ \ {d}_{n+1}\geqslant{d}_{n},
\end{equation}
which guarantees that the void state remains isolated during the time
between state preparation and outcome detection.

In general, a non-zero signal labstate $|\psi,n)$ will be a linear
combination of terms of the form $\mathbb{A}_{i_{1},n}^{+}\mathbb{A}%
_{i_{2},n}^{+}\ldots\mathbb{A}_{i_{k},n}^{+}|0,n)$, for some integer $k>0$
and with $i_{1}<i_{2}\ldots<i_{k}.$ Consider Schr\"{o}dinger evolution from
time $n$ to time $n+1$. If we write
\begin{equation}
\mathbb{A}_{i_{1},n}^{+}\mathbb{A}_{i_{2},n}^{+}\ldots\mathbb{A}%
_{i_{k},n}^{+}|0,n)\rightarrow\mathbb{U}_{n+1,n}\mathbb{A}_{i_{1},n}^{+}%
\mathbb{A}_{i_{2},n}^{+}\ldots\mathbb{A}_{i_{k},n}^{+}|0,n)
\end{equation}
then using closed semi-unitarity we find%
\begin{align}
\mathbb{U}_{n+1,n}\mathbb{A}_{i_{1},n}^{+}\mathbb{A}_{i_{2},n}^{+}\ldots%
\mathbb{A}_{i_{k},n}^{+}|0,n) & =\left\{ \mathbb{U}_{n+1,n}\mathbb{A}%
_{i_{1},n}^{+}\mathbb{U}_{n+1,n}^{+}\right\} \left\{ \mathbb{U}_{n+1,n}%
\mathbb{A}_{i_{2},n}^{+}\mathbb{U}_{n+1,n}^{+}\right\} \ldots  \notag \\
& \left\{ \mathbb{U}_{n+1,n}\mathbb{A}_{i_{k},n}^{+}\mathbb{U}%
_{n+1,n}^{+}\right\} |0,n+1),
\end{align}
which means that, in principle, a knowledge of the transitions
\begin{equation}
\mathbb{A}_{i,n}^{+}\rightarrow\mathbb{U}_{n+1,n}\mathbb{A}_{i,n}^{+}\mathbb{%
U}_{n+1,n}^{+}
\end{equation}
of the individual signal operators should give us complete knowledge about
the dynamics.

In general, a given signal operator may evolve to a multiple-signal
operator. Such a scenario may occur when experiments are performed on bound
states, for example. An initially prepared bound state would be described by
a one-signal labstate which could then evolve into a two or more-signal
labstates. That would correspond to an experiment detecting particle decay
products.

\section{Path summation}

Given a normalized initial labstate
\begin{equation}
|\psi,M)\equiv\sum_{i=0}^{{d}_{M}-1}\psi^{i}|i,M)
\end{equation}
at time $M,$ then semi-unitary evolution gives%
\begin{equation}
|\psi,M)\rightarrow\mathbb{U}_{M+1,M}|\psi,M)\equiv\sum_{j=0}^{{d}%
_{M+1}-1}\sum_{i=0}^{{d}_{M}-1}U_{M+1,M}^{j,i}\psi^{i}|j,M+1),
\end{equation}
where the coefficients $\left\{ U_{M+1,M}^{j,i}\right\} $ satisfy
semi-unitarity, provided $r_{n+1}\geqslant r_{n}$. Then the amplitude $%
\mathcal{A}(j,M+1|\psi,M)$ to go to $|j,M+1)$ from the initial labstate $%
|\psi,M)$ is given by
\begin{equation}
\mathcal{A}(j,M+1|\psi,M)=\sum_{i=0}^{{d}_{M}-1}U_{M+1,M}^{j,i}\psi^{i}.
\end{equation}
Semi-unitarity then leads to total probability conservation, i.e.,
\begin{equation}
\sum_{j=0}^{{d}_{M+1}-1}|\mathcal{A}(j,M+1|\psi,M)|^{2}=1.
\end{equation}

Suppose however that at time $M+1,$ the observer does not attempt to
determine any outcome but in effect channels the labstate into new apparatus
(to use the language of SQM). Then $|\psi,M+1)$ serves as an initial state
for the next jump, and so on. After a sequence of successive semi-unitary
jumps, ending at time $N$, the amplitude $\mathcal{A}(j,N|\psi,M)$ to go to $%
|j,N)$ for $N>M$ is given by
\begin{equation}
\mathcal{A}(j,N|\psi,M)=\sum_{i=0}^{{d}_{N-1}-1}U_{N,N-1}^{j,i}\mathcal{A}%
(i,N-1|\psi,M).  \label{111}
\end{equation}
It is easy to see that total probability is conserved in this case. We note
that semi-unitarity requires $r_{M}\leqslant r_{M+1}\leqslant\ldots\leqslant
r_{N}$. The particular scenario where strict equality occurs corresponds to
what happens in SQM, where it is usual to identify successive Hilbert spaces
with each other. In such a case, semi-unitarity can be replaced by
unitarity, so that the dynamics appears reversible.

Expression (\ref{111}) can be written out in the form of a path-summation,
i.e.,%
\begin{equation}
\mathcal{A}(j,N|\psi ,M)=\sum_{i_{N-1}=0}^{{d}_{N-1}-1}\sum_{i_{N-2}=0}^{{d}%
_{N-2}-1}\ldots \sum_{i_{M}=0}^{{d}%
_{M}-1}U_{N,N-1}^{j,i_{N-1}}U_{N-1,N-2}^{i_{N-1},i_{N-2}}\ldots
U_{M+1,M}^{i_{M+1},i_{M}}\psi ^{i_{M}},
\end{equation}%
which is the SFQM version of the Feynman path integral. We note that in the
Feynman path integral, as it is conventionally formulated over physical
space, there is an implicit assumption that the particle could be observed
in principle anywhere in physical space. Since space is regarded as a
continuum in SQM, the formalism naturally leads to integration rather than
summation. Additionally, the time parameter is taken to a continuum limit,
which is a source of severe technical problems in SQM. Neither of these
continuity assumptions are made in the SFQM approach.

\subsection{Time reversal experiments}

It is possible, under carefully controlled circumstances, to violate the
semi-unitarity inequality, i.e., to have an experiment where $%
r_{n+2}<r_{n+1}>r_{n}$ whilst maintaining total probability conservation and
a good physical interpretation. Such a possibility occurs in time reversal
experiments, for instance, where the observer would carefully arrange their
apparatus in such a way so as to ensure $%
U_{n+2,n+1}^{j,i}=(U_{n+1,n}^{i,j})^{\ast }$. Magnetic resonance experiments
are specifically designed to test the degree to which this relationship
holds in the presence of temperature-dependent (irreversible) processes.

\section{Concluding remarks}

We have presented a general framework for encoding quantum principles in
instrumentalist terms, consistent with Bohr-Heisenberg reality. By focusing
on what is physically meaningful, i.e., the apparatus, and not on any
supposed system under investigation, we find a formalism which deals with
quantum physics in a realistic way. The focus now is on information
acquisition and loss, which is all that experimentalists ever deal with.

We do not have the space here to discuss a number of important
and related issues, such as the role of null experiments in
creating a \textquotedblleft multi-fingered \textquotedblright
view of time, and the dynamical generation of spacelike and
timelike causal structures in quantum processes. Neither do we
have room to discuss various specific applications
to quantum physics, such as quantum optics networks
\cite{JAROSZKIEWICZ-2006}, which generally confirm the validity
and usefulness of SFQM.

\end{document}